\input{aipcheck}

\documentclass[
    ,final            
  ]
  {aipproc}

\layoutstyle{6x9}

\begin{document}

\title{ UNEXPECTED GOINGS-ON IN THE STRUCTURE OF A NEUTRON STAR CRUST   }

\author{Aurel Bulgac}{
  address={Department of Physics, University of Washington, Seattle WA
    98195-1560, USA}
}

\author{Paul-Henri Heenen}{
  address={Service de Physique Nucl\'eaire Th'eorique, Univerite Libre de
    Bruxelles,B 1050, Brussels, Belgium} 
}

\author{Piotr Magierski}{
  address={Faculty of Physics, Warsaw University of Technology, ul. Koszykowa
    75, 00-662, Warsaw, Poland }
}

\author{Andreas Wirzba}{
  address={Helmholtz-Institut f\"ur Strahlen- und Kernphysik,
Universit\"at Bonn,D-53115 Bonn, Germany}
}
\author{Yongle Yu}{
            address={Department of Physics, University of Washington, Seattle WA
    98195-1560, USA}
}
\begin{abstract}

We present a brief account of two phenomena taking place in a neutron star
crust: the Fermionic Casimir effect and the major density depletion of the 
cores of the superfluid neutron vortices. 

\end{abstract}

\maketitle


\section{Fermionic Casimir effect and Neutron Star Crust }

At a depth of about 500 m or so below the surface of a neutron crust the
nuclear matter (which consists mostly of neutrons plus a small percentage of
protons and electrons in $\beta$-equilibrium) organize themselves in some
exotic inhomogeneous solid phase \cite{bbp}. As a matter of fact, neutron star
crusts seem to be just about the only other places in the entire Universe,
apart from planets, where one can find condensed matter, in particular a solid
phase \cite{pwa}. Moving from the neutron star surface inward, one finds at
first a Coulomb crystal lattice of nuclei immersed in a very low density neutron 
gas and even lower density electron gas. With increasing depth, the density and
pressure increase, the nuclei get closer to each other and start evolving into
some unusual elongated nuclei, which eventually become rods. These nuclear rods
evolve gradually into plates, their place being taken later by tubes and
bubbles (dubbed ``inside out'' nuclei) just before the average density becomes 
almost equal to the nuclear saturation density and the entire mixture of
neutrons, protons and electrons become an homogeneous phase. The properties of
this part of the neutron star have been the subject of a lot of studies, see Refs. 
\cite{bbp,nvb,bv,pra,dhm,dha,mh,mbh,bma1,bw,bma3,bma4,bma5,jap1,pethick,jones} and
other references therein. Most of these approaches however have missed a rather
subtle and apparently important physical phenomenon, the fermionic counterpart
of the Casimir interaction in such a medium \cite{bma1,bw,bma3,bma4,bma5}. 

In order to quickly explain the main physics ideas behind this new phenomenon,
let us consider an over-simplified model of the neutron star crust. One can ask
the rather innocuous question: ``What is the ground state energy of an infinite
homogeneous Fermi sea of noninteracting neutral particles with two hard spheres of radii
$a$, separated by a distance $r$?'' The naive and somewhat startling answer
that perhaps one can place the two hard spheres almost anywhere with respect to
each other and that the energy of the system will not be affected if one were
to move the hard spheres around. The ``theoretical argument'' which can lead to
such a conclusion is based on the same type of argumentation, which was used in
Refs.\cite{bbp,pra,dhm,dha} and allowed these authors to establish that by going
deeper and deeper into the interior of the neutron star one finds a well
defined sequence of ``exotic'' nuclear shapes. This traditional argumentation
is based essentially on liquid drop model, which includes the volume, surface,
Coulomb contributions to the ground state energy only. This is basically
``classical thinking.'' For a person using ``quantum reasoning'' instead, the
fact that the ground state of such a system in infinitely degenerate
(corresponding to an arbitrary relative arrangement of the two hard spheres)
will find such an answer most likely wrong. An indeed, a careful analysis of
the problem reveals the fact that indeed a system of two hard spheres, immersed
in an infinite Fermi see of noninteracting particles at zero temperature has a
well defined ground state. The correct answer, namely that the ``interaction
energy'' of the two hard spheres of radius $R$, at distance $r$ from each other, is 
somewaht even more surprising. One finds that
$$
E_C \approx -\frac{\hbar^2k_F^2}{m} \frac{R^2}{2\pi r(r-2R)} j_1[2k_F(r-2R)],  
$$
where $j_1(x)$ is the spherical Bessel function, $k_F$ is the Fermi momentum
and $m$ is the fermion mass. ``Why would this ``interaction energy'' be a
non-monotonic function of the hard sphere separation $r$?'' and, moreover, ``How
does interaction really emerges here, where one starts with such a simple
system of non-interacting particles?'' As one soon ``discovers'' the
``culprit'' is the wave character of the Quantum Mechanics really. Fermions
even at zero temperature do not stop moving and the space is really ``filled''
with an infinite number of de Broglie's waves. These waves reflect from the two
hard spheres and as in the case of any wind musical instrument, for some
frequencies one would have a favorable wave interference while for other
frequencies there will not such a favorable interference. In an infinite Fermi
sea there is an infinite number of waves with all frequencies ranging from zero
to the Fermi frequency. If one carefully adds up the effects of all these
waves one readily arrives at the result above \cite{bma1,bw}. 
Things get a little bit more complicated when one adds more hard spheres, as
then one naturally discovers that besides the ``natural'' two-body interactions
there are genuine three- and four- and many-body interactions among these
spheres. Moreover, there is absolutely no reason why not consider other type of
objects, which could be immersed in this Fermi sea, like ``logs'' and
``boards'' and in principle almost anything else. Surprisingly all these
combinations of various objects in various arrangements can be analyzed rather
easily. What is surprising however is the fact that the characteristic
interaction energy between such objects is of the same order as the energy
differences between various phases in a neutron star crust
\cite{bma1,bma3,bma4,bma5} and when taken into account this fermionic Casimir
energy can in ``ruin perfect crystalline structures'' found in all previous
studies. These conclusions have been backed by more sophisticated fully
microscopic calculations of the nuclear matter in a neutron star crust
\cite{mh,mbh}. 

Instead of describing in more detail results which have been published already,
we shall instead draw the attention of our readers here to another element
which was overlooked in studies of the neutron star crust, and which is
apparently going to influence a great deal of properties. 
In order to analyze the thermal and electric conductivities
of the crust, which are important for understanding of the thermal
evolution of neutron stars one has to go beyond the static approximation. The
``nuclei'' which are immersed in the neutron fluid, which indeed is a
superfluid, can and do move. As with boats on a lake, when they start moving
they make waves and one has to include the dynamics of the surrounding
superfluid in any analysis. We shall limit ourselves here to quoting a single
result, namely the kinetic energy of two penetrable spheres,located at the distance r,
 immersed in a
superfluid at velocities below the critical velocity for the loss of
superfluidity. One then finds \cite{mb03} that the kinetic energy of two such
spheres becomes:
$$
T=  \frac{1}{2}(M_{1}^{ren}u_{1}^{2}+M^{ren}_{2}u_{2}^{2} )  \\
 + 4\pi m\rho_{out} \left (\frac{1-\gamma}{2\gamma+1}\right )^{2}
                      \left (\frac{R_{1}R_{2}}{r}\right )^{3}\left [
\vec{u}_{1}\cdot\vec{u}_{2}-
\frac{3}{r^{2}}(\vec{u}_{1}\cdot\vec{r})(\vec{u}_{2}\cdot\vec{r})
\right ]
$$
where the renormalized masses of nuclei have the form:
$$
M_{i}^{ren}=\frac{4}{3}m\rho_{in}\pi R_{i}^3
\frac{(1-\gamma)^{2}}{2\gamma+1}=
            M_{i}\frac{(1-\gamma)^{2}}{2\gamma+1},
$$
where $\vec{u}_i$ are the velocities of the two nuclei,
$i=1,2$ and $M_{i}$ and $R_i$ denote the nuclear bare mass and radii of the $i-th$
nucleus, $\gamma = \rho_{out}/\rho_{in}$ and $\rho_{in,out}$ are the densities
inside and outside the two nuclei. The somewhat unexpected cross term appearing
above shows that the existence of mere motion of the two objects in a perfect
fluid can lead to a velocity-dependent interaction, which decays with the
separation as slows as the static Casimir Fermionic energy, namely as $1/r^3$.
Further analysis shows that this velocity dependent-interaction is important as
well when considering dynamical properties of neutron star crust \cite{pm}.

\section{The Spatial Structure of a Vortex in Low-Density Superfluid Neutron Matter  }

There is a long held belief that vortices in Fermi systems do not show any
appreciable normal density variations and that only the anomalous density
vanishes along the vortex axis, similarly to the behavior of the density (which
is the order parameter) in Bose systems \cite{gennes,aads,deblasio}. Thus it
came as somewhat of a surprise the fact that in Fermi systems one can have a
spatial structure of a vortex with a significant normal density depletion along
the vortex axis \cite{vortex,pierre,abyy}. What happens in low density
superfluid neutron matter for example is the following. The magnitude of the
pairing gap becomes comparable with the Fermi energy, 

The possibility that the value of the superfluid gap can attain large
values was raised more than two decades ago in connection with the BCS
$\rightarrow$ BEC crossover \cite{leggett,nozieres}. One can imagine
that one can increase the strength of the two--particle interaction in
such a manner that at some point a real two--bound state forms, and in
that case $a\rightarrow -\infty$. By continuing to increase the
strength of the two--particle interaction, the scattering length
becomes positive and starts decreasing.
A dilute system of fermions, when $\rho r_0^3\ll 1$ (here $r_0$ is the
interaction radius), will thus undergo a
transition from a weakly coupled BCS system, when $a<0$ and $a
={\cal{O}}(r_0)$, to a BEC system of tightly bound Fermion pairs, when
$a>0$ and $a ={\cal{O}}(r_0)$ again. In the weakly coupled BCS limit
the size of the Cooper pair is given by the so called coherence length
$\xi \propto \frac{\hbar^2k_F}{m\Delta},$
which is much larger than the inter-particle separation $\approx
\lambda_F=2\pi/ k_F$. In the opposite limit, when $k_Fa\ll 1$ and
$a>0$, and when tightly bound pairs/dimers of size $a$ are formed, the
dimers are widely separated from one another. Surprisingly, these
dimers also repel each other with an estimated scattering length
$\approx 0.6...2a$ \cite{mohit,pieri} and thus the BEC phase is also
(meta)stable.
 The bulk of the theoretical analysis in the intermediate
region where $k_F|a|>1$ was based on the BCS formalism
\cite{mohit,leggett,nozieres,randeria} and thus is highly
questionable. Even the simplest polarization corrections have
not been included into this type of analysis so far. In particular, it
is well known that in the low density region, where $a<0$ and
$k_F|a|\ll 1$ the polarization corrections to the BCS equations lead
to a noticeable reduction of the gap \cite{gorkov}.  Only a truly {\it
ab initio} calculation could really describe the structure of a many
Fermion system with $k_F|a|\gg 1$. In the limit $a=\pm \infty$, when
the two--body bound state has exactly zero energy, and if $k_Fr_0\ll
1$, one can expect that the energy per particle of the system is
proportional to $\varepsilon_F=\hbar^2k_F^2/2m$, as it was recently
confirmed by the variational calculations of
Refs. \cite{carlson1,carlson2}.
The normal density at the vortex core is lowered, while the pairing field
vanishes at the vortex axis as expected.
In hindsight this
result could have been expected. Large values of the pairing field
correspond to the formation of atom pairs/dimers of relatively small
sizes. When these dimers are relatively strongly bound and when they
are also widely separated from one another, they undergo a
Bose--Einstein condensation. For a vortex state in a 100\% BEC system
the density at the vortex axis vanishes identically. Therefore, by
increasing the strength of the two--particle interaction, the Fermion
system simply approaches more and more an ideal BEC system, for which
a density depletion of the vortex core is expected.

Almost thirty years ago Anderson and Itoh \cite{pwa} put forward the
idea that vortices should appear in neutron stars and that they can
also get pinned to the solid crust. They argued that the
"star--quakes," observable on Earth as pulsar "glitches," apparently
 are caused by the vortex de--pinning in neutron star crust. This idea and its
various implications have been examined by numerous authors, see
Refs. \cite{aads,pierre} and further references therein, but a general
consensus does not seem to have emerged so far. 

The profile of a vortex in neutron matter
is typically determined using a Ginzburg--Landau equation, which is
expected to give mostly a qualitative picture and its accuracy is
difficult to estimate.  Surprisingly, prior to Ref. \cite{vortex} there exists
only one microscopic calculation of a vortex in low density neutron matter
\cite{deblasio}.   
The existence of a strong density depletion in the vortex core is
going to affect appreciably the energetics of a neutron star crust.
One can obtain a gross estimate of the pinning energy of
a vortex on a nucleus as $E^V_{pin}=[\varepsilon(\rho_{out}) \rho_{out} -
\varepsilon(\rho_{in}) \rho_{in}]V$, where $\varepsilon(\rho)$ is the energy per
particle at density $\rho$, $\rho_{in}$ and
$\rho_{out}$ are the densities inside and outside the vortex core and
$V$ is the volume of the nucleus. Naturally, this simple formula does
not take into account a number of factors, in particular surface
effects and the changes in the velocity profile and the pairing
field. These last contributions were accounted for (with some
variations) in the past \cite{pwa,aads}. However, if the density
inside the vortex core and outside differ significantly one expects
$E^V_{pin}$ to be the dominant contribution.  In the low density
region, where $\varepsilon(\rho_{out})\rho_{out}/\varepsilon(\rho_{in})\rho_{in}$
is largest,  one expects a particularly large
anti--pinning effect $(E^V_{pin}>0)$. The energy per unit length of a
simple vortex is expected to be significantly lowered when compared
with previous estimates \cite{pwa,aads} by $\approx [\varepsilon(\rho_{out})
\rho_{out} - \varepsilon(\rho_{in}) \rho_{in}]\pi R^2$, where $R$ is an
approximate core radius.

\end{document}